# Active Extensile Hydrogels Actuated by Living Polymers of the Bacterial Cytokinetic Protein FtsZ

Mikheil Kharbedia[1,2], Diego Herráez-Aguilar[1,3], Macarena Calero[1,2], Horacio López-Menendez[1,2], Clara Luque-Rioja[1,2], Lara H. Moleiro[1,2], Cruz Santos[3], Pilar Lillo[4], and Francisco Monroy[1,2,*].

[1] *Department of Physical Chemistry, Universidad Complutense, Ciudad Universitaria 28040 Madrid (Spain)*
[2] *Unit of Translational Biophysics, Instituto de Investigación Hospital Doce de Octubre (i+12), Av. Córdoba 28041 Madrid (Spain)*
[3] *Instituto de Investigaciones Biosanitarias, Universidad Francisco de Vitoria, Ctra. M-515 Pozuelo-Majadahonda Km 1,800, Pozuelo de Alarcón, Madrid, Spain*
[4] *Instituto de Química Física "Blas Cabrera", C/ Serrano 119, 28006 Madrid (Spain)*

**\* e-mail: monroy@ucm.es**

## ABSTRACT

Active materials capable of autonomously modulating their mechanical properties are foundational to the development of next-generation soft technologies. Here, we introduce a novel class of extensile biohybrid hydrogels powered by living polymers of the bacterial cytokinetic protein FtsZ. When embedded within a polyacrylamide (PA) matrix, GTP-fueled FtsZ filaments self-organize into treadmilling structures that generate internal extensile stresses, driving reversible softening, swelling and fluidization of the composite FtsZ–PA hydrogel network. Unlike conventional contractile biopolymer systems, these hybrid gels exhibit stress-induced softening, yield under minimal deformation, and suppress thermal flow barriers—hallmarks of dissipative, extensile metamaterials. Microscopic particle tracking reveals active non-Gaussian fluctuations, while bulk rheology confirms programmable, concentration-dependent reductions in both stiffness and viscosity. Theoretical modeling shows that internal filament activity gives rise to a negative mechanical permittivity, establishing a new paradigm in materials science in which embedded FtsZ living polymers dynamically program active soft matter mechanics from within. These findings open new avenues for the design of modular, reconfigurable systems in adaptive biomaterials, soft robotics and synthetic active matter.

# INTRODUCTION

Nature excels at building active force generators from living biopolymers[1–4] —*energy-consuming filaments powered by nucleotide hydrolysis*— actuating across biomolecular scales[5–8]. Inspired by this paradigm, we introduce a novel class of extensile biohybrid materials: mechano-active hydrogels powered by living polymers of FtsZ, a cytoskeletal protein universally conserved in bacteria[9–11]. FtsZ self-assembles into GTP-dependent, treadmilling filaments that form the cytokinetic Z-ring during cell division[12,13]. Functionally, treadmilling FtsZ serves both as a dynamic scaffold for septation[14], and as a force generator driving membrane constriction[15,16]. Its "living polymerization" is reversible, $Mg^{2+}$-dependent, and occurs above a critical concentration of ~0.1 mg/mL in vitro[17]. Each monomer hydrolyzes one GTP upon turnover, producing a living polymer capable of continuous remodeling[17-21]. Reconstituted FtsZ filaments act autonomously in biomimetic systems[22-25], confirming their potent mechanical activity. While the precise mechanism of cytokinetic force generation remains debated[17-28], these results identify FtsZ filaments as promising autonomous internal actuators for synthetic soft matter devices. Building on this foundation, we develop an actively extensile hydrogel by embedding randomly treadmilling FtsZ filaments into a compliant polyacrylamide (PA) matrix. Upon GTP activation, filament activity induces distributed extensile stresses that modulate mechanical properties throughout the gel. These multidimensional effects scale with filament concentration[29]—a hallmark of active matter behavior[30–31]. Multiscale modeling of FtsZ–PA composites confirms that this mechanical cooperation arises from activity-induced modulation of neo-Hookean elasticity under suppression of constrained reptation dynamics[29]. This establishes FtsZ-based gels as a versatile model system for exploring active soft materials driven by living polymers, a direction emphasized in recent active matter roadmaps[32].

Functionality-driven advanced hybrid hydrogels[33-37] adapt their viscoelastic properties in response to mechanical stress and environmental signals (temperature, osmotic pressure, pH)[38], opening avenues for applications ranging from biosensing to drug delivery. Their hydrated, compliant networks provide an ideal platform for embedding bioactive components, enabling the design of hybrid materials with emergent properties distinct from the passive matrix[33-38]. Among such bioactives, mechano-responsive biomolecules—including shape-altering proteins[39,40], molecular motors[41-43] and living cytoskeletal filaments[44,45], can act as mechanical primers converting chemical energy into localized, non-thermal forces[Error! Marcador no definido.,38]. When incorporated at low concentrations to avoid network self-percolation or mechanical crowding, such biomolecular actuators can generate spatially distributed active stresses within the hydrogel—provided the host matrix remains sufficiently soft to accommodate these perturbations. Active hydrogels incorporating living biofilaments are categorized as either *contractile,* where embedded force-generating actuators (e.g., actomyosin[42] or microtubule-kinesin[43] systems) exert pulling forces to compact the material, or *extensile,* where internal filaments generate pushing forces that expand and soften the matrix. While contractile gels have been widely demonstrated[1-8], biologically based extensile gels remain comparatively underexplored, particularly in systems where internal stress generation emerges from isotropic, stochastic filament dynamics. Beyond its specific biochemical components, the FtsZ–polyacrylamide hybrid presented here serves as a prototype for a broader class of internally active composite hydrogels, where force-generating agents—biomolecular, synthetic, or engineered—can be embedded within soft elastic matrices to program frequency-dependent mechanical responses. The theoretical framework we develop, combining Soft Glassy Rheology (SGR)[46], with Landau-type bifurcation mechanics[29], provides a general blueprint for designing tunable viscoelastic metamaterials driven by internal activity, irrespective of the microscopic actuator.

In this work, we harness FtsZ—*the prokaryotic motor protein recognized for its dynamic self-assembly into reconfigurable living filaments*[20], to fabricate novel hybrid hydrogels exhibiting actively extensile capacity driven by polymerization-mediated forces. Embedding these randomly oriented, GTP-fueled FtsZ filaments within a compliant polyacrylamide matrix creates a unique composite material where filament treadmilling generates internal stresses that soften and fluidize the host, yielding emergent extensibility. Contrasting with classical contractile systems (actomyosin[41,42], tubulin filaments[8] and microtubule-kinesin networks[43]), FtsZ-driven hybrid PA-hydrogels display active softening akin to *hyperfluidity,* and act as activatable metamaterials with negative mechanical permittivity: *amplifying applied stress rather than resisting it.* In a prior work[29], we established a theoretical framework for understanding the active forces generated within hybrid FtsZ-based hydrogels, emphasizing the amplification role of hydrodynamic modes of FtsZ treadmilling leading to active reptation across the passive matrix. Here, we extend that non-equilibrium framework to account for the nonlinear weakening responses observed under increasing strain and deformation frequency. This non-equilibrium mechanism of extensile actuation, fundamentally distinct from equilibrium matrix resistance, opens a new avenue for designing programmable, soft active gel materials based in living-polymers. This advance enables the future integration of living polymer systems into adaptive biomedical scaffolds, self-healing materials, and energy-transforming components for soft robotics. By inverting the typical resistant hydrogel response to deformation, actively compliant FtsZ-based systems offer a powerful new approach to metamaterials design, exhibiting potential applications ranging from tissue engineering to soft robotics. Upon realizing a unique form of extensile actuation, biohybrid FtsZ-hydrogels expand the design metamaterial toolbox of mechanically programmable devices, with promising applications in drug delivery[47], cell-instructive regenerative scaffolds[48], and novel soft functionalities using polymer-based biointegrated engineering strategies[49,50].

## RESULTS

**Living FtsZ Polymers Induce Non-Equilibrium Fluctuations.**

**Figure 1** provides direct experimental evidence for the mechanical activity of FtsZ filaments, using high-speed videomicroscopy of probe microparticles embedded in FtsZ-containing solutions. In the absence of GTP, the medium contains only monomeric FtsZ, and particle motion appears confined with displacements well-described by a Gaussian distribution i.e., a particle thermally trapped in a harmonic potential (Fig. 1A). Upon GTP addition, filaments polymerize reversibly under the catalytic presence of $Mg^{2+}$, rendering the system active. Strikingly, despite the expected increase in structural stiffness due to filament formation, particles exhibit enhanced motility with broader, non-Gaussian displacement distributions characterized by pronounced heavy tails (Fig. 1B). The diffusive trajectories reveal a clear transition from passive confinement to active propulsion. In the passive state, the motion is consistent with an Ornstein–Uhlenbeck process, $MSD_{pass}(t) \approx 4Dt_{conf}(1 - e^{-t/t_{conf}}) \approx 4Dt$, where the confinement time is approximately $t_{conf} \approx 10\ s$. Upon GTP activation, the motion becomes hyperdiffusive, characterized by a ballistic-like MSD scaling: $MSD_{act}(t) \approx c_F^2 t^2$, with a fitted propulsion velocity $c_F = (2.7 \pm 0.3)\ \mu m/s$ (Fig. 1, right panel). This dynamic transition reflects a shift in the viscoelastic medium from passive thermal equilibrium to an active, non-equilibrium fluctuating state driven by GTP consumption. In the passive state, the displacement probability distribution, $P(x) = exp(-U_{conf}/k_B T)$, corresponds to a harmonic confinement potential, $U_{conf}^{(pass)} = k_{pass} x^2/2$, where $k_{pass}$ represents the effective stiffness by the surrounding medium. Upon activation by GTP, this confinement potential is significantly modified. Empirically, the active distribution is consistent with a softened, asymmetric potential: $U_{conf}^{(act)} \approx k_{act} x^2/2 - F_0|x|$, where $k_{act}$ ($< k_{pass}$) reflects a decrease in rigidity by the polymerizing FtsZ filaments, and $F_0$ quantifies the average active force exerted by them. In our analysis, GTP-induced filament activation led to a measurable decrease in stiffness, from $k_{pass} = (15.3 \pm 1.8)\ pN/\mu m$ down to $k_{act} = (10 \pm 3)\ pN/\mu m$, which is accompanied by a characteristic active force of $F_0 \approx 50 pN$, notably higher than the force typically exerted by actomyosin strokes[41,42]. This active softening force internally accounts for the observed long-tailed, non-Gaussian displacement statistics (Fig. 1B). A force stroke of $F_0 \approx 50 pN$ acting over a short burst time $t \ll t_{conf} \approx 10\ s$, (e.g., milliseconds) imparts kinetic energy, $E_K = F_0 \Lambda_0$, with characteristic displacements, $\Lambda_0 \approx 100\ nm$, much longer and energetic than merely thermal fluctuations. This active energy injection, corresponding to $E_K \approx 5 \times 10^{-18}\ J \approx 10^3\ k_B T$, far exceeds thermal energy and suffices to drive the particle out of its harmonic trap, resulting in active bursts that punctuate the otherwise confined dynamics. These power strokes represent an active stress source embedded within the viscoelastic network and exemplify the capacity of living FtsZ polymers to induce stochastic mechanical transitions at the mesoscale. Together, these findings confirm that FtsZ filaments function as internal mechanical actuators, simultaneously softening the matrix and injecting localized stress.

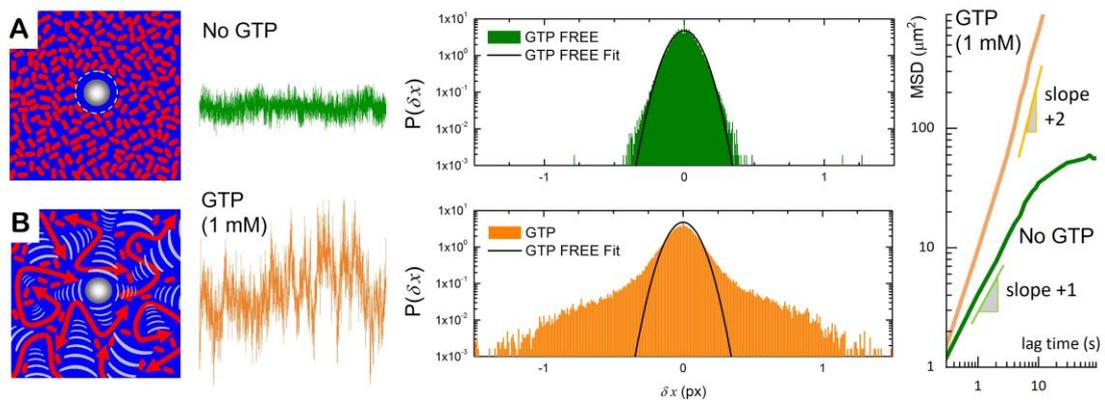

**Figure 1. Mechanical activity of FtsZ filaments.** The viscoelastic medium induces Brownian motion in embedded microparticles (1 μm diameter). A probe microparticle (gray) embedded in the active network undergoes Brownian motion, as shown in the corresponding time traces captured at high temporal resolution: **A) Passive case (no GTP):** In the absence of GTP, FtsZ remains in its monomeric state and no internal activity is present (schematic, left). The corresponding time series shows confined thermal fluctuations with a high thermal stochasticity, and the displacement distribution (green) is well fit by a Gaussian profile, consistent with harmonic trapping in a viscoelastic medium. **B) Active case (+GTP):** Upon GTP addition (1 mM), FtsZ filaments dynamically polymerize and treadmill, driving active fluctuations in the medium (schematic, left). This results in markedly enhanced particle displacements exhibiting high volatility and a very broad, heavy-tailed (non-Gaussian) distribution (orange), indicative of out-of-equilibrium dynamics and internal force generation. <u>Rightmost panels:</u> Mean-square displacement (MSD) analysis shows that particles transition from confined subdiffusion (passive state; $MSD \sim t$) to hyperdiffusive motion in the active regime ($MSD \sim t^2$). The crossover time (~10 s) reflects the confinement timescale of the passive network. The observed shift in the confinement potential and the emergence of active, asymmetric forces validate the role of FtsZ filaments as extensile internal actuators in the hydrogel matrix. All measurements were performed using a 40× objective, where 1 pixel corresponds to 120 nm, allowing accurate resolution of submicron displacements across time.

The emergent active (non-equilibrium) dynamics and anomalous particle Brownian motion are signatures of active gel behavior, substantiating the extensile mechanical role of living FtsZ polymers.

**Structural Remodeling of Polyacrylamide Gels by Embedded FtsZ Filaments.**

To asses the microstructural impact of embedding FtsZ filaments into polyacrylamide (PA) hydrogels, we performed transmission electron microscopy (TEM) imaging on samples with two crosslinking densities: 2.5% and 7%, corresponding to low and high rigidity states. **Figure 2** presents TEM images of the microstructures formed within the synthetic hydrogels under the experimental conditions. In each case, control gels lacking FtsZ were compared with their FtsZ-containing counterparts, prepared at concentrations exceeding the critical polymerization threshold for filament formation (Fig. 2, left and right panels, respectively). At 2.5% crosslinking (Fig. 2A), the control PA gels exhibited large, well-defined porous domains with a polygonal architecture (Fig. 2A, left), indicative of low-density, open-mesh structures. Upon incorporation of polymerizing FtsZ (Fig. 2A, right), the matrix showed substantial architectural disruption: the regular geometry was replaced by a more irregular, fibrous network with evident signs of internal wrinkling, collapse and anisotropic annealing of the polymer mesh. This structural transformation suggests a local reorganization of the hydrogen network driven by internal FtsZ activity, consistent with internal stress propagation and active remodeling. At 7% crosslinking (Fig. 2B), the matrix displayed a much denser and more isotropic morphology, with small, interlinked pores and a relatively uniform fibrous texture in the absence of FtsZ (Fig. 2B, left). Interestingly, when FtsZ filaments were embedded at this higher crosslinking density (Fig. 2B, right), the effect was notably subtler. Although minor rearrangements and densification were observed, the tightly crosslinked network largely retained its original structure, suggesting that mechanical feedback from FtsZ activity is less effective in stiffer matrices where the elastic constraints resist filament-induced deformation. Together, these observations underscore the mechanically active role of embedded FtsZ filaments and their capacity to structurally remodel soft hydrogel matrices—most prominently at lower crosslinking densities, where the polymer network remains sufficiently compliant. The emergent anisotropy and collapse patterns observed via TEM provide direct visual evidence of internal actuation and softening effects mediated by living polymer activity within the hybrid material (see Fig. 2; right inset image for schematic microstructural depiction).

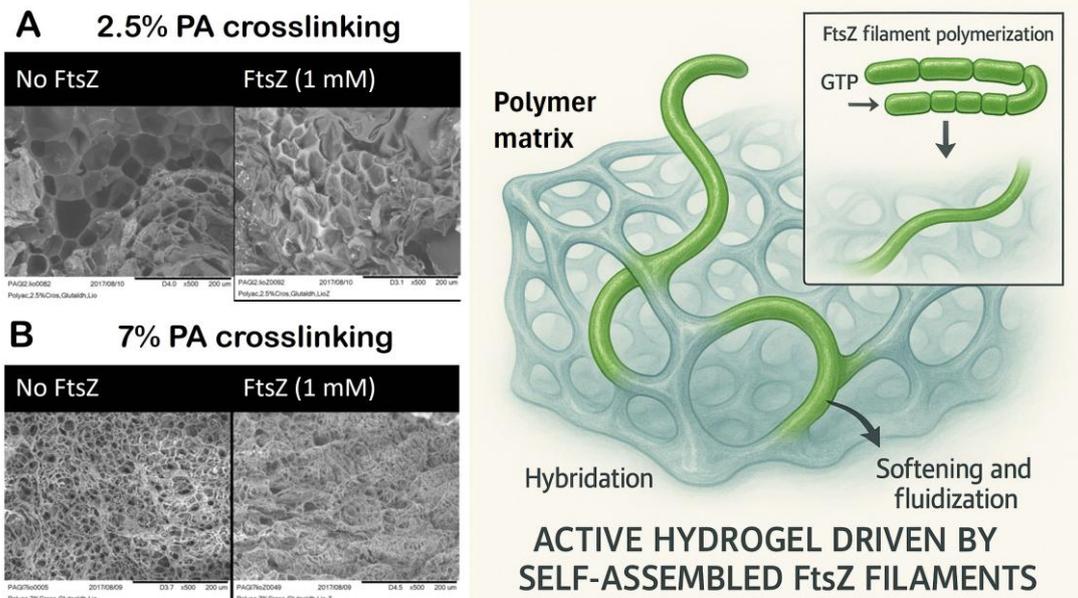

**Figure 2. Ultrastructural changes in FtsZ-polyacrylamide hybrid hydrogels visualized by TEM:** Transmission electron microscopy images of hydrogels with 2.5% (top panels) and 7% (bottom panels) crosslinking, in the absence (left) and presence (right) of polymerizing FtsZ filaments. <u>Top left:</u> Passive hydrogel (2.5% crosslinking) shows polygonal, ordered pore structure typical of a low-crosslinked polyacrylamide network. <u>Top right</u>: Upon addition of FtsZ at concentrations above the critical polymerization threshold, the same gel exhibits disrupted pore symmetry and corrugated matrix features, indicating filament-induced reorganization. <u>Bottom left:</u> At higher crosslinking density (7%), the passive gel reveals a dense, sponge-like porous texture. <u>Bottom right:</u> Inclusion of polymerized FtsZ induces clear matrix remodeling, resulting in a less uniform, more fibrillar architecture. These structural differences underscore the mechanical influence of GTP-fueled FtsZ dynamics on the hydrogel architecture, supporting the extensile remodeling inferred from rheological and morphometric analyses. **Inset)** Cartoon depiction of an active hydrogel system in which FtsZ filaments, polymerizing in the presence of GTP, integrate within a passive polyacrylamide (PA) matrix (image generated by GPT 4o DALL-E). The hybridization of the dynamic filaments with the polymer network drives internal stress generation, leading to local matrix softening and fluidization. This biomimetic design illustrates how active, self-assembling cytoskeletal elements can impart tunable mechanical behavior to synthetic hydrogels.

**Structural Extensibility and Mechanical Softening.** To probe the extensibility of hybrid FtsZ–polyacrylamide (PA) hydrogels under internal activation, we prepared microgel droplets using a millifluidic setup comprising a needle-terminated injection line (see Methods). The precursor solution consisted of 3% acrylamide, 0.07% bis-acrylamide, 0.5 mg/mL FtsZ, and 1 mM GTP, but excluded $Mg^{2+}$ to prevent premature filament polymerization. Individual droplets (~100 μm diameter) were formed in an isotonic aqueous environment containing initiator agents, allowed to equilibrate overnight under osmotic and mechanical conditions. Following equilibration, the droplets were transferred into isotonic buffer solutions either without (control) or with 2 mM $Mg^{2+}$ to catalyze GTP hydrolysis and activate FtsZ polymerization. As shown in **Figure 3 (left panel)**, droplets incubated without $Mg^{2+}$ retained their original shape and size (cyan contours), whereas $Mg^{2+}$-exposed specimens underwent rapid and significant radial expansion (magenta contours), reaching a ~42% average volume increase within 5 minutes of activation. This inflation response was strictly GTP-dependent and fully reversible. After ~6 hours, the droplets gradually deflated, returning close to their initial volume as GTP was hydrolyzed and FtsZ filaments depolymerized. The final gel morphology exhibited slight irregularities consistent with spatially heterogeneous filament disassembly. Quantitative morphometric analysis of volume changes (**Figure 3, right panel**) confirmed a robust and reproducible inflation in the activated condition (mean $\Delta V/V = 42\%$, $N = 13$, $\sigma = 5.3\%$), while control samples remained essentially unchanged ($\Delta V/V \approx 0\%$, $N = 15$, $\sigma = 1.5\%$). These observations establish that GTP-activated FtsZ polymerization within the PA matrix generates internal extensile forces capable of deforming the surrounding gel. The reversible swelling behavior observed in these FtsZ-PA microdroplets supports the interpretation that the hybrid hydrogel network exhibits structural extensibility accompanied by transient mechanical softening, driven by the dynamic activity of embedded living FtsZ filaments.

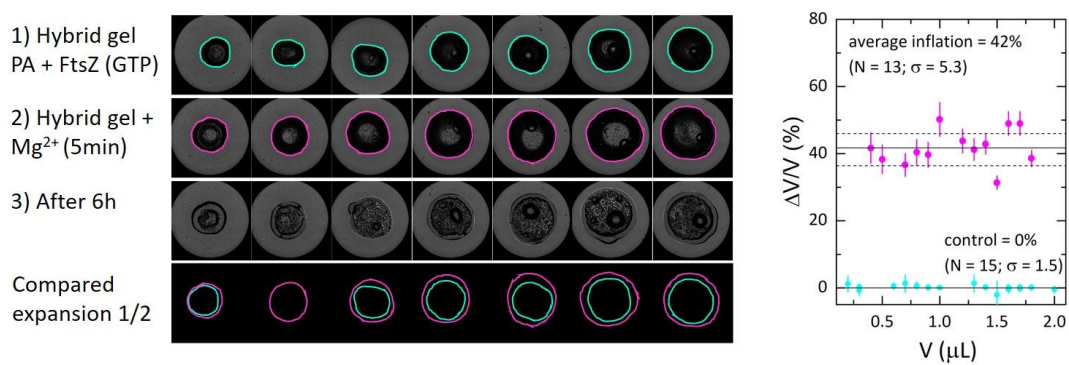

**Figure 3. Structural extensibility of hybrid FtsZ–polyacrylamide (PA) microspheres upon GTP-dependent filament activation.** *Left strips' panel:* Time-lapse optical microscopy of individual hydrogel microspheres shows volumetric inflation upon FtsZ polymerization. *1) Initial state:* hybrid FtsZ–PA microspheres formed with GTP but without $Mg^{2+}$ remain unchanged. *2) After 5 min of $Mg^{2+}$ addition (2 mM),* polymerization of FtsZ is triggered, leading to visible expansion of the droplets. *3) After 6 h,* droplets partially recover their initial volume as GTP is hydrolyzed and filaments depolymerize. *Bottom strip:* superimposed contours of the same microspheres before and after activation highlight the extent of expansion. *Right panel:* Morphometric analysis of microdroplet volume change (ΔV/V) as a function of initial volume (where N is the number of specimens analyzed, and σ is the standard deviation). Active specimens (magenta) show an average inflation of 42%, while control samples (cyan, no $Mg^{2+}$) remain stable (0% change). These results confirm the reversible, internal stress-induced extensibility of the hybrid gel, consistent with mechanical softening driven by living FtsZ polymer activity.

**Active FtsZ polymerization: mechanical softening and fluidization.** To directly assess the dynamic mechanical response of hybrid hydrogels under biochemical activation, we monitored the shear modulus (G') and viscosity (η) of macroscopic specimens in real time during controlled GTP-induced FtsZ polymerization. As shown in **Figure 4**, the hybrid gel composed of PA hydrogel and dispersed FtsZ monomers exhibits baseline mechanical stiffening in the absence of GTP, consistent with progressive matrix crosslinking. However, upon the addition of $Mg^{2+}$—a necessary cofactor for GTP hydrolysis and filament formation—a marked, sudden drop in both G′ and η is observed, defining a clear transition to an actively softened and hyperfluidized state (**Fig. 4A**, yellow left panels). This mechanical transition reflects the onset of internal stress generation by the treadmilling FtsZ filaments, which act as extensile actuators disrupting the hydrogel network. In sharp contrast, control formulations lacking either GTP or FtsZ monomers do not show this behavior (**Fig. 4B**, gray right panels). These samples maintain their initial rigidity or exhibit only minor softening attributable to passive swelling. The specific, GTP-dependent onset of softening therefore confirms that the mechanical transition is driven by filament activity and not by osmotic or swelling-related effects alone. These findings represent the first quantitative demonstration of an active extensile gel system whose viscoelastic properties are modulated from within by mechanically dynamic living FtsZ biopolymers. The abrupt mechanical response underscores the ability of FtsZ filaments to inject localized forces into the gel matrix, effectively tuning its macroscopic rheology in a reversible and programmable manner.

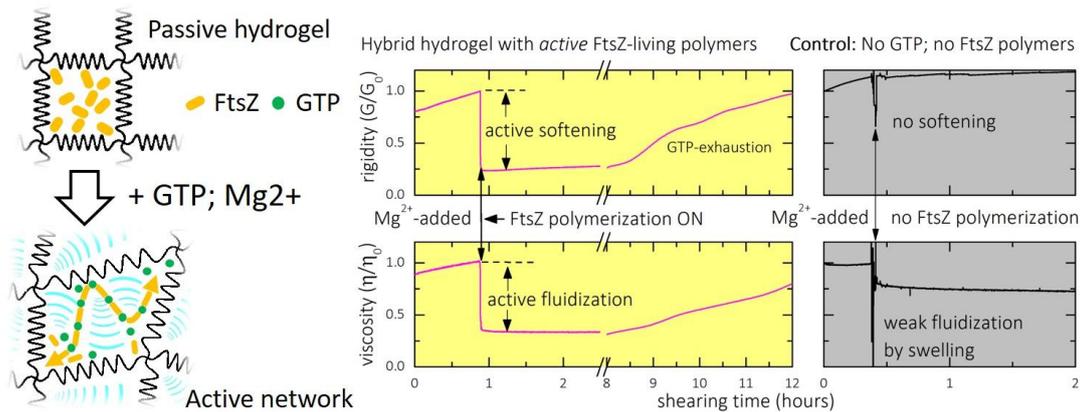

**Figure 4. Active softening and fluidization of FtsZ-embedded hydrogels.** Schematic (top) shows the transition from a passive polyacrylamide hydrogel with dispersed FtsZ monomers (left) to an active extensile network upon GTP- and $Mg^{2+}$-induced filament polymerization (right). Rheological measurements reveal that in the active system (panel A, yellow background), FtsZ polymerization causes a sudden drop in both rigidity (G') and viscosity (η), indicating active softening and fluidization. This softening persists until GTP exhaustion after 8 hours. In contrast, the control condition lacking GTP (panel B, gray background) shows no softening and only minor fluidization due to passive swelling. These results confirm that active FtsZ filaments dynamically reduce matrix resistance via internal extensile stress, establishing the material as an actively swellable, reconfigurable hydrogel.

**Network Composition Effects in Active FtsZ–Hydrogels: Elasticity Mapping.** To evaluate the structural mechanics and extensile capacity of FtsZ-containing hydrogels, we performed a systematic rheological analysis across varying polymer concentrations and filament activity states. **Figure 5** consolidates these data, combining compositional dependencies and temporal responses to highlight the unique extensile behavior of active biohybrid networks under shear deformation. Specimen photographs at the top strip illustrate the sol–gel transition of passive hydrogels in inclined capillaries, with structural rigidity appearing sharply above 2.5% acrylamide. This sol-gel transition is rheologically captured in **Figure 5A,** which plots the terminal shear modulus G' across concentration for three hydrogel formulations: 1) bare polyacrylamide (PA) gel (black); 2) passive FtsZ-blended PA hydrogel (cyan, without polymerization) ; 3) active FtsZ-polymerized hydrogels (magenta, with GTP and Mg$^{2+}$). Scaling analysis reveals a critical percolation threshold near $c_{crit} \approx 2.5\%$, with power-law behavior consistent with classical gelation theory by De Gennes[51], this is $G' \approx G_0(c - c_{crit})^3$ at $c > c_{crit}$. Crucially, in all cases, the active gels display lower G' values than the pasive ones, confirming internal softening via extensile stress generated by filament activity. **Figure 5B** shows the time evolution of G' under constant oscillatory shear. Passive matrices undergo slow stiffening due to residual crosslinking, while GTP-free FtsZ blends remain mechanically stable. In contrast, active gels—after $Mg^{2+}$-induced polymerization, undergo abrupt softening, confirming that FtsZ filaments, when activated, reduce the gel's rigidity in real time. This softening is absent in control formulations, underscoring its dependence on the biochemical activation of FtsZ. **Figure 5C** extends this analysis by quantifying the terminal G' as a function of FtsZ concentration for three acrylamide densities (3%, 4%, and 5%). A monotonic, dose-dependent decrease in G' is observed, with the most pronounced effect near the percolation threshold. This supports the hypothesis that softening is amplified in less crosslinked networks, where extensile forces more easily propagate. Photographic evidence provides a macroscopic view of the sol–gel transition (see tilting essays in Fig. 5). Passive PA hydrogels (top strip) retain their shape at acrylamide concentrations above ~2%, consistent with the onset of network percolation and gelation. In contrast, active hybrid gels containing FtsZ at concentrations ≥2 mg/mL (bottom strip) exhibit persistent shape collapse across the same PA range. This behavior reveals a state of macroscopic hyperfluidity, where externally induced deformation leads to flow, despite the presence of an underlying polymer network, highlighting the dominant fluidizing effect of self-assembled FtsZ filaments. These visual cues correlate directly with rheological data, offering clear mechanical confirmation of active softening. Collectively, these results establish a quantitative framework for designing FtsZ-based active gels with tunable extensile mechanics and programmable rheological responses via filament concentration and network density. This underscores the potential of filament-based living polymers to endow soft materials with dynamically reconfigurable mechanical properties.

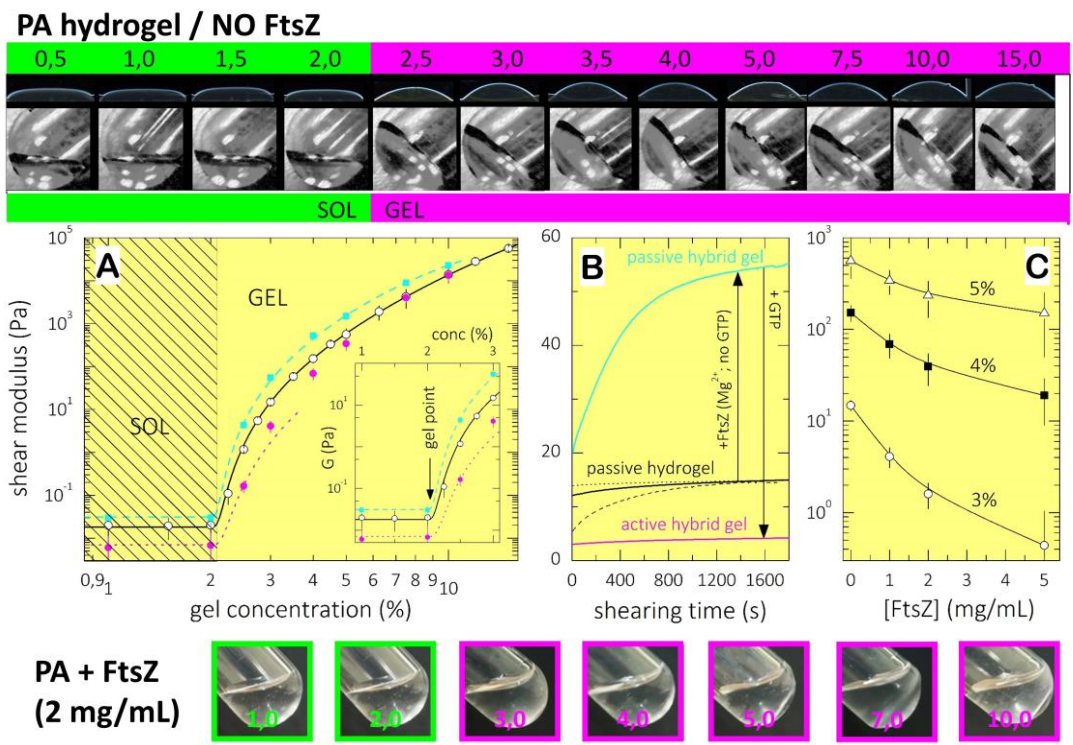

**Figure 5. Composition-dependent rheology of passive and active FtsZ–polyacrylamide (PA) hydrogels.** *Top strip:* Tilting tests showing the sol–gel transition across PA concentrations (0.5–15%). Pure PA hydrogels (no FtsZ) transition from liquid-like sol (green, <2.5%) to solid-like gel (magenta, >2.5%), retaining their shape upon tilting. **A) Sol-gel rheological transition:** Shear modulus (G′) as a function of gel concentration for passive PA hydrogel (black), passive hybrid gels with FtsZ but no GTP (cyan), and active hybrid gels with GTP-polymerized FtsZ (magenta). The inset (log–log scaling) highlights a percolation-like transition near ~2.5% concentration. Active hybrids show consistent softening relative to passive gels, confirming extensile remodeling by living FtsZ filaments. **B) Hybrid filamentation process:** Time-dependent rheology under oscillatory shear shows dynamic stiffening in passive gels, including non-polymerized FtsZ blends. Upon GTP-triggered polymerization (arrow), active hybrids exhibit abrupt softening, reflecting internally driven fluidization of the matrix. **C) Active fluidization under increasing living FtsZ polymerization:** Terminal shear modulus (G′) as a function of FtsZ concentration for three PA densities (3%, 4%, 5%). All exhibit monotonic softening with increasing FtsZ concentration, most pronounced in the lowest-density gels, consistent with active disruption of the percolated network. *Bottom strip:* Optical tilt tests of hybrid hydrogels with fixed FtsZ (2 mg/mL) and varying PA content (1–10% ; expressed in box) confirm persistent liquid-like behavior (green frame) throughout the entire PA concentration range in the gel regime (cyan frame). This contrasts with the solid appearance of pure PA gels above the sol–gel threshold (green to cyan color code), emphasizing the dominant active role of embedded FtsZ filaments in a hyperfluidized meshwork.

**Rheological signatures in hybrid FtsZ–PA hydrogels.** To dissect the non-equilibrium mechanical response imparted by FtsZ activity, we performed systematic rheological measurements on three types of hydrogels: (i) passive polyacrylamide (PA) gels, (ii) passive gels doped with FtsZ monomers (no GTP), and (iii) active hybrid gels in which FtsZ filaments are dynamically polymerized in the presence of GTP and Mg$^{2+}$. These systems were studied across increasing hydrogel density (3%, 5%, and 10%, at fixed crosslink concentration 7.5%), which modulate the baseline elasticity of the polymer network. **Figure 6A** displays stress–strain curves under oscillatory shear, revealing distinct mechanical signatures across the three conditions. At low strains, all systems exhibit a linear (Hookean) regime consistent with Newtonian-like behavior. However, as strain increases, yield points become evident—marking the onset of irreversible deformation. These yield points shift to lower strain values in the active gels, particularly at low polymer density (3%), indicating reduced structural integrity due to internal activity. Notably, the active gels transition directly from the linear regime into a plastic flow regime with no strain-stiffening, in sharp contrast to the passive and monomeric controls, which show a more gradual yielding transition. This mechanical softening upon deformation confirms the internal disruption of the network by extensile activity of treadmilling FtsZ filaments. **Figure 6B** quantifies the **f**requency-dependent viscoelastic response through the storage modulus $G'$, and loss modulus $G''$, measured via small-amplitude oscillatory shear at 37 °C. At all crosslinker concentrations, passive gels—with or without monomeric FtsZ—exhibit classic elastic solid-like behavior, with $G' > G''$ and minimal frequency dependence of $G'(\sim\omega^0)$. These rheological features are consistent with a stable crosslinked polymer network. The loss modulus ($G''$), follows a weak power-law scaling $G''\sim\omega^{0.25}$, ndicative of slow, glassy dissipation characteristic of the soft glassy rheology (SGR) regime[46]. In contrast, active hybrid gels show a marked softening in both moduli. The storage modulus $G'$ is significantly reduced, and the loss modulus $G''$ increases with a steeper slope, up to $G''\sim\omega^{0.75}$ especially at low crosslinking (3%). This steeper scaling suggests the presence of sub-Maxwellian fluidity, where energy injected by active treadmilling drives the system away from equilibrium, leading to enhanced viscous dissipation and reduced elastic storage. Importantly, even in this active, fluidized state, the rheological response does not fully conform to a Newtonian fluid: rather, it reflects a non-equilibrium viscoelastic state with memory, consistent with active extensions of the SGR framework. Overall, these results support a dynamic crossover from solid-like to fluid-like behavior modulated by both network density and internal activity. As polymerr concentration increases (left to right), all gels become more rigid, but the active gels remain significantly softer, illustrating how internal polymerization forces override structural reinforcement. This interplay between passive architecture and active disruption underlies the mechanical tunability of hybrid cytoskeletal gels, relevant to both synthetic active matter and cytoplasmic mechanics.

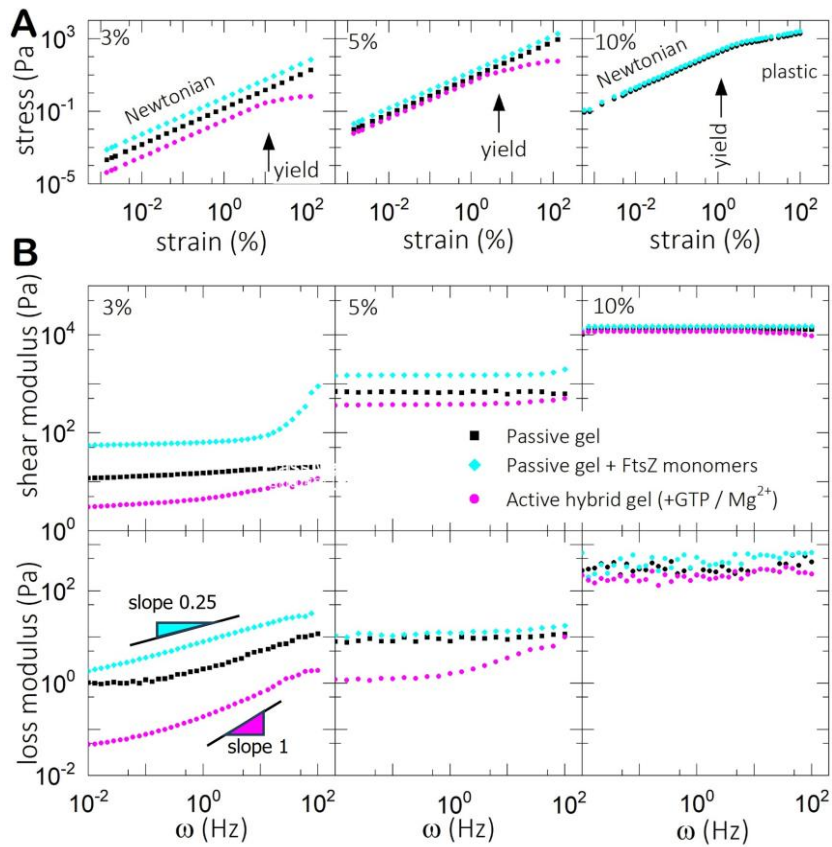

**Figure 6. Rheological characterization of passive and active FtsZ–polyacrylamide hydrogels. A)** Stress–strain curves for hydrogels of increasing polymer crosslinking (3%, 5%, and 10%) under oscillatory shear, showing deviations from Newtonian behavior. Passive (black) and monomeric FtsZ-blended gels (cyan) exhibit linear viscoelasticity followed by yielding and strain stiffening. In contrast, active gels containing GTP-polymerized FtsZ filaments (magenta) display nonlinear softening and, at low gel densities, plasticity-like behavior, indicating loss of mechanical integrity due to internal filament activity. **B)** Frequency-dependent viscoelastic moduli measured at 37 °C. Top: shear modulus (G′); bottom: loss modulus (G″). Data are shown for three PA concentrations (3%, 5%, and 10%). Active gels (magenta) exhibit pronounced frequency-dependent softening, especially at low crosslinking (3%), while passive and monomer-only systems maintain frequency-invariant behavior. The power-law scaling of G′ and G″ in active conditions is consistent with soft glassy rheology, where internal activity modulates the material's mechanical noise temperature.

**Mechanical permittivity: non-reciprocal susceptibility in active gels.** The mechanical behavior of FtsZ–PA hybrid hydrogels arises from the interplay of internal force generation and nonlinear elasticity[29], resulting in non-reciprocal mechanical responses. These materials can be conceptualized as mechanically polarizable media, where internal FtsZ filaments act as dipolar stress generators, analogous to electric dipoles in dielectric metamaterials[52]. GTP-fueled treadmilling drives internal activity that modulates the material's effective stiffness, generating active stresses that either reinforce or oppose externally applied deformations[29]. At the macroscopic continuum scale, this is formalized as a non-reciprocal stress balance:

$$\sigma_{ij}^{(eff)} = \sigma_{ij}^{(ext)} - \sigma_{ij}^{(act)} \tag{1a}$$

where $\sigma_{ij}^{(ext)}$ is the externally applied stress and $\sigma_{ij}^{(act)}$ is the internally generated active component. The constitutive law follows a nonlinear elastic form:

$$\sigma_{ij} = \mathbf{G}(u_{ij})u_{ij} \tag{2}$$

where $u_{ij}$ is the strain tensor and $\mathbf{G}(u_{ij})$ is a strain-dependent stiffness, consistent with a neo-Hookean response (see SM in Ref. 29). In passive systems, $\mathbf{G} = \mathbf{G}_0$, is a strain-dependent stiffness, consistent with a neo-Hookean response (see SM in Ref. $\sigma_{ij}^{(ext)} = \mathbf{G}_0 u_{ij}$, with $\mathbf{G}_0$ denoting the baseline matrix rigidity. In active gels, internal activity modulates the effective stiffness as $\mathbf{G}_{eff}(u_{ij}) = \mathbf{G}_0 \pm \Delta\mathbf{G}$, leading to:

$$\sigma_{ij}^{(act)} = \sigma_{ij}^{(eff)} - \sigma_{ij}^{(ext)} = (\mathbf{G}_{eff} - \mathbf{G}_0)u_{ij} = \pm\Delta\mathbf{G}\, u_{ij}, \tag{3}$$

Here, $\pm\Delta\mathbf{G} \equiv \mathbf{G}_{eff} - \mathbf{G}_0$ represents the signed excess stiffness due to internal activity. A positive $+\Delta\mathbf{G}$ corresponds to contractile stiffening ($\sigma_{act} > 0$), whereas a negative $-\Delta\mathbf{G}$ reflects active extensile softening ($\sigma_{act} < 0$). This motivates the definition of a mechanical permittivity[38]:

$$\epsilon_{\pm} = \mathbf{G}_{eff}/\mathbf{G}_0 = 1 \pm \Delta\mathbf{G}/\mathbf{G}_0 = 1 \pm \chi, \tag{4}$$

where $\chi \equiv tr(\Delta \mathbf{G} \cdot \mathbf{G}_0^{-1})$ is a mechanical susceptibility quantifying the relative modulation of stiffness. This framework mirrors dielectric theory[52], with the stress relation in Eq. (1a) rewritten as (see Supplementary Note N1 for full derivation and bifurcation structure):

$$\sigma_{ij}^{(eff)} = \epsilon_\pm \sigma_{ij}^{(ext)} = \sigma_{ij}^{(ext)} \pm \chi \sigma_{ij}^{(ext)} = (1 \pm \chi)\sigma_{ij}^{(ext)} \qquad (1b)$$

Analogous to dielectric polarization, $P \approx \epsilon_0 \chi E$, with the full displacement, $D = \epsilon_0 E + P(E) = \epsilon_0(1 + \chi)E$, here the effective mechanical displacement under the external polarization field is[52]:

$$\sigma_{ij}^{(eff)} = \sigma_{ij}^{(ext)} + \sigma_{ij}^{(int)} \quad with \quad \sigma_{ij}^{(int)} = \pm \chi \sigma_{ij}^{(ext)} \qquad (5)$$

This formalism gives rise to non-reciprocal stress propagation, where the permittivity $\epsilon_\pm = 1 \pm \chi$, encodes the internal reactivity to applied strain.

**Figure 7A** schematically illustrates the bifurcation structure of the stress–strain response governed by mechanical susceptibility $\chi$. The passive reference state corresponds to zero susceptibility ($\chi = 0$), yielding a linear elastic response ($\epsilon = 1$, gray line). Positive susceptibility ($\chi > 0$), corresponds to contractile activity, resulting in active stiffening ($\epsilon_+ > 1$), where internal stress reinforces the applied deformation. Conversely, negative susceptibility ($\chi < 0$) reflects extensile activity that reduces the effective stiffness ($0 < \epsilon_- < 1$), corresponding to stable softening (orange curve). At large strains or high activity, the system may cross into an unstable regime ($\epsilon_- < 0$), where internally generated stress opposes external work, leading to negative mechanical impedance and potential mechanical collapse. This bifurcation structure captures the onset of non-reciprocal polarization induced by internal filament dynamics and defines the boundaries of mechanical stability (see Supplementary Note N1, Eqs. S6–S9, for the analytical derivation and bifurcation conditions). **Figure 7B** quantifies the relative rigidity modulation, defined as $\chi = \Delta G_0'/G_0' = (G_{FtsZ}' - G_0')/G_0'$ across different gel concentrations and frequencies. In passive gels, increasing polymer content reinforces stiffness ($\chi > 0$), as expected from elastic percolation. In contrast, active gels exhibit pronounced softening, particularly within the low-density range (3–5%), where the matrix is sufficiently compliant to be reorganized by internal FtsZ-driven stress generation. This elastic softening, indicative of negative susceptibility ($\chi < 0$), reflects active extensility and confirms that internal FtsZ forces dominate the elastic response in compliant networks. **Figure 7C** shows the corresponding relative loss modulation, $\varphi = \Delta G_0'' / G_0'' = (G_{FtsZ}'' - G_0'')/G_0''$, as a function of gel concentration. Passive gels display increased viscous losses with increasing polymer density ($\varphi > 0$), consistent with standard viscoelastic buildup. Active gels, however, exhibit a systematic decrease in loss modulus, signaling a transition to a more fluidized regime under extensile stress ($\varphi < 0$).

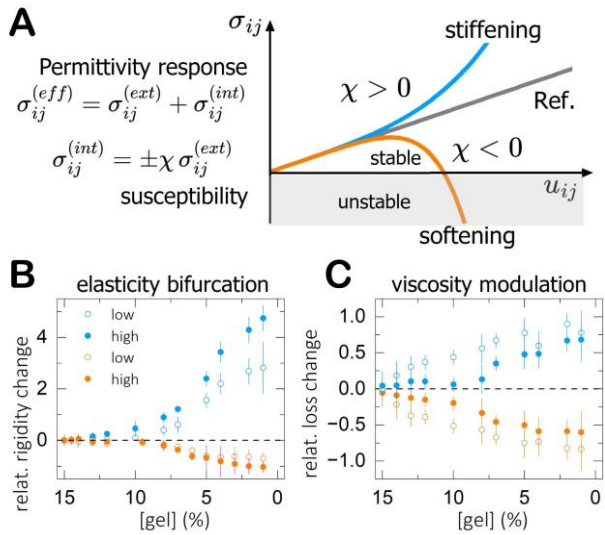

**Figure 7. Mechanical bifurcation in passive and active FtsZ–polyacrylamide hydrogels** (constant FtsZ concentration: 2 mg/mL). **A)** Schematic representation of the mechanical bifurcation induced by internal FtsZ activity, formulated through a mechanical permittivity model. The total effective stress $\sigma_{ij}^{(eff)}$ esults from the sum of externally applied stress $\sigma_{ij}^{(ext)}$, and internally generated activity-driven stress $\sigma_{ij}^{(int)} = \pm \chi \sigma_{ij}^{(ext)}$, where $\chi$ is the frequency-dependent mechanical susceptibility. Positive susceptibility ($\chi > 0$) leads to active stiffening, while negative susceptibility ($\chi < 0$) leads to extensile softening, bifurcating from the passive reference line ($\chi = 0$, black). Depending on the magnitude of activity and strain, the softening branch can exist either within the absolutely stable regime ($\sigma, d\sigma > 0$), or in a marginally stable domain ($\sigma > 0, d\sigma < 0$), where the system approaches mechanical instability ($\sigma < 0$; grey dashed region). **B)** Relative rigidity modulation, defined as the normalized change in storage modulus ($\chi = \Delta G'/G_0'$) plotted as a function of gel concentration for low and high frequency regimes. As an elasticity modulation, passive gels (blue) exhibit increased stiffness with crosslink density, while active gels (orange) show pronounced softening, especially in low-density networks (3–5%). Frequency-dependent effects are evident, with larger susceptibility (greater $\Delta G'$) observed at high frequencies. **C)** Relative losses modulation, given by the normalized change in viscous dissipation ($\varphi = \Delta G''/G_0''$), also shows maximal softening at low gel concentrations. Active gels consistently display reduced loss moduli compared to passive controls, consistent with activity-driven fluidization and energy injection. These results support a bifurcation framework based on mechanical permittivity and susceptibility. Error bars denote standard deviations across independent measurements.

This trend, maximal again at 3–5% gel concentration, highlights the role of filament activity in reducing dissipation through structural reorganization. These results confirm the emergence of a concentration-dependent bifurcation regime of mechanical susceptibility, where internal activity governs macroscopic rheology.

Moreover, both properties ($\chi$ and $\varphi$) reveal systematic frequency-dependent differences. At high frequency ($\omega = 10^2 \, Hz$), both the rigidity modulation ($\chi = \Delta G_0'/G_0'$), and the loss modulation ($\varphi = \Delta G_0''/G_0''$), are larger in magnitude than at low frequency. This trend is particularly evident in active gels, where internal FtsZ activity couples more efficiently to faster strain rates, enhancing the dynamic contribution of internal stress generation. At low frequencies ($\omega = 10^{-2} \, Hz$), the active filaments have time to reorganize and adapt to imposed deformations, partially dissipating active FtsZ stress via remodeling. In contrast, at high frequencies, the rapid deformation outpaces internal reorganization, leading to a reactive buildup of internal stress and, consequently, an amplified modulus modulation by FtsZ polymers. These results are consistent with the expectation $\epsilon(\omega) = 1 \pm \chi(\omega)$, supporting the interpretation of active gels as frequency-tunable mechanical metamaterials. Their stiffness, dissipation, and even the sign of internal stress can be modulated through both composition and strain rate (see Supplementary Note N1, Eq. S15–S19, for frequency-dependent stability criteria and active modulus modulation thresholds). In passive systems ($\chi = 0$), hence $\epsilon = 1$, so that no internal reactive stress arises: $\sigma_{ij}^{(int)} = 0$. In contrast, active systems exhibit $\sigma_{ij}^{(int)} \neq 0$ ($\chi \neq 1$), with $\epsilon_+ > 1$ for contractile ($\chi > 0$), and $\epsilon_- < 1$ for extensile ($\chi < 0$) activity. The onset of bulk extensility corresponds to a positive trace, $tr(\Delta \mathbf{G}) < 0$, indicating a negative modulus correction—a hallmark of gain-like behavior in active matter[29]. Experimentally, this manifests as GTP-induced gel expansion (Fig. 3) and load-induced softening (Figs. 4–6), both indicative of negative mechanical impedance[29]. Taken together, these results identify a mechanically programmable regime in FtsZ–PA hydrogels, defined by active bifurcation of rigidity and loss under internal stress generation. This behavior directly supports the Landau-type phase transition model (described in Supplementary Note N1), where $\chi(\Gamma)$ emerges as a secondary order parameter, directly linked to internal activity and strain, modulated by a primary structural order parameter, $\Gamma$. This theoretical framework rationalizes the transition between stiffened, passively elastic networks and softened, actively extensible materials with frequency-dependent rheological signatures.

**Metamaterial FtsZ-driven permittivity: active viscoelastic compliance.** To quantify the activity-induced modulation of mechanical impedance, we measured the **complex mechanical permittivity**, defined as $\tilde{\epsilon}(u,\omega) = \tilde{G}/\tilde{G}_0$, which represents the frequency- and strain-dependent mechanical compliance of the FtsZ–PA hybrid gel, normalized to its passive polyacrylamide baseline. This dimensionless rheological response serves as a mechanical analog to dielectric permittivity in generative media[52], capturing how internal FtsZ activity modifies both elastic and dissipative properties of the network (see Supplementary Note N1, for analytical model of complex permittivity in the SGR–Landau framework). **Figure 8A** shows the frequency dependence of the real part, $\epsilon'(\omega)$, which reflects the elastic compliance. In all passive gels (3%, 5%, and 10% PA), high permittivity ($\epsilon' = 1 + \chi > 1$), is observed consistently across frequencies, indicating contractile stiffening behavior imposed by polyacrylamide crosslinking. This trend reflects the dominant contribution of the PA matrix in the absence of internal activity. In contrast, active gels containing treadmilling FtsZ exhibit a strong suppression of real permittivity ($\epsilon' < 1$), particularly at higher frequencies and in less crosslinked networks (3% PA). This active regime corresponds to a negative susceptibility ($\chi < 0$), meaning the material exhibits a reactive response to deformation that leads to extensility rather than contractility. Here, filament turnover actively injects energy into the network, generating internal stress fluctuations that suppress elastic resistance. This leads to extensile softening, where applied deformation is amplified by the system's internal dynamics rather than opposed. The effect is most pronounced in low-stiffness networks and at high frequencies, where the viscoelastic matrix becomes increasingly responsive to internally driven, non-equilibrium force fluctuations from active FtsZ treadmilling. In this regime, elasticity is no longer governed solely by the passive matrix, but instead emerges from a dynamic balance between structural resistance and active softening forces, captured quantitatively by a negative rheological susceptibility ($\chi_{act} < 0$, thus $\epsilon'_{act} < 1$). **Figure 8B** presents the frequency-resolved fluidity parameter: $\phi = \omega/\epsilon''$, which captures the system's susceptibility to viscous deformation. All gel compositions display increasing fluidity with frequency, as expected, but active gels exhibit a dramatically steeper slope in the bilogarithmic plot. This steepening is indicative of enhanced internal actuation, driven by persistent nonequilibrium fluctuations from FtsZ treadmilling. Notably, active samples exceed ($\phi \approx 10^2 - 10^3$), at frequencies above 10 Hz, identifying a transition into an active mechanical state where internal energy injection dominates over viscoelastic damping. This active-fluid regime is strongly suppressed in highly crosslinked (10%) or passive samples, suggesting that matrix rigidity constrains internal rearrangement, effectively filtering active stresses and preventing fluidization. The observed trends in both $\epsilon'$ and $\phi$ are remarkably well reproduced by the analytical expression derived in the SGR–Landau model (see Supplementary Note N2, Eqs. S29–S31). This model captures the crossover from passive to active regimes via a tunable control parameter connected to susceptibility $\chi(x)$, encoding both rheological noise ($x$), and mechanical symmetry breaking ($\Gamma$). The model fits yield excellent agreement with experimental data across all PA concentrations (see best fitting parameters in **Table I, Supplementary Note N2**), validating the interpretation of $\chi$ as an activity-dependent susceptibility and confirming the relevance of the SGRL framework. Notably, the susceptibility interpreted as an order parameter from the Landau theory, correlates with the effective noise temperature inferred from the SGR model (see **Figure S1**, **Supplementary Note N2**). This quantitative agreement provides compelling support for the SGRL permittivity as a predictive descriptor of active matter rheology. It captures not only the elastic suppression and fluidization trends but also enables a mechanistic decomposition of internal dissipation and stiffness in terms of underlying biochemical and structural dynamics.

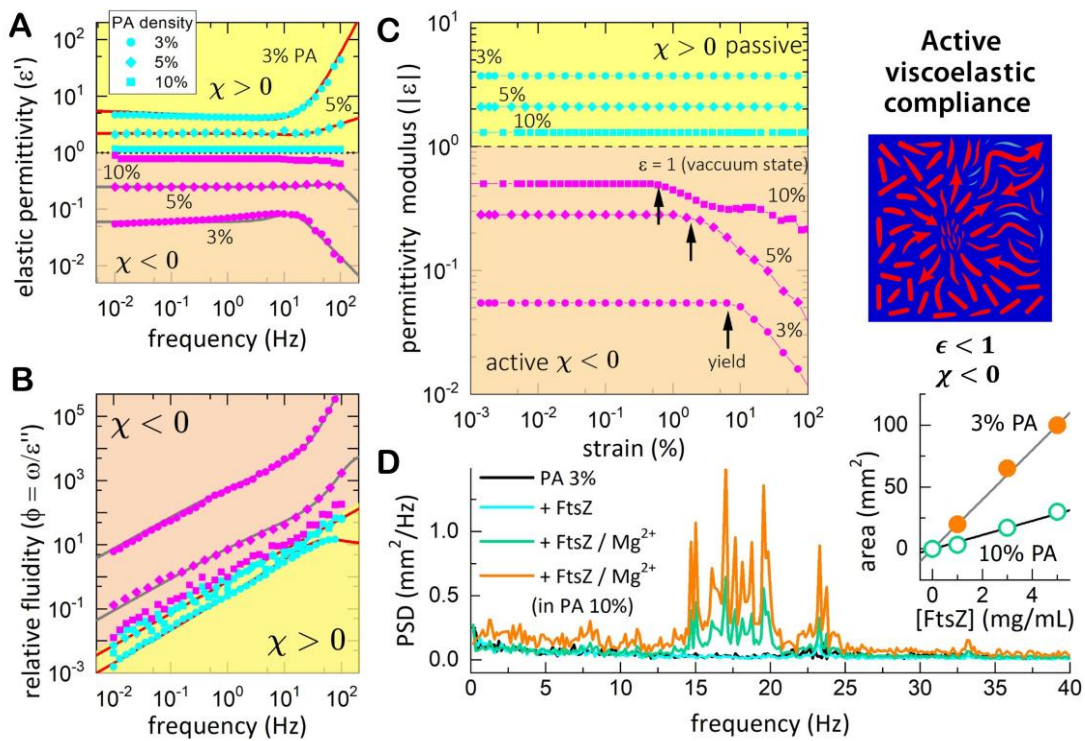

**Figure 8. Viscoelastic response of hybrid FtsZ-PA hydrogel material under internal force generation. A) Elastic permittivity** ($\epsilon'$) as a function of frequency shows a plateau at low frequencies and a frequency-dependent regime above 1 Hz, modulated by activity level for constant FtsZ concentration (2 mg/mL) at different PA density (symbols: 3%, 5%, 10%; straight lines: best fittings to Eq. 7). **B) Relative fluidity** ($\phi = \omega/\epsilon''$) increases with frequency, indicating enhanced dissipative response at higher activity (symbols as in A: 3%, 5%, 10%; straight lines: best fittings to Eq. 8; see main text for discussion). **C) Permittivity modulus** ($\epsilon$) as a function of strain highlights a transition from a passive, linear regime above glass ($\epsilon > 1$) to an active, strain-softening regime ($\epsilon < 1$). The yield point shifts to higher strain with increasing activity. *Right:* schematic of active viscoelastic compliance depicting force-induced rearrangement in active filament networks. The central source of treadmilling activity illustrates regions of sub-unitary permittivity ($\epsilon < 1$), characteristic of active softening under compliance ($\chi<0$). **D) Laser Doppler velocimetry (LDV) spectra of hybrid hydrogels.** Distinct FtsZ hydrodynamic modes of treadmilling emerge upon activation of FtsZ polymerization in PA hydrogels (7% density; 2.5 % crosslinking) containing FtsZ (5 mg/ml) under $MgCl_2$ addition (20 mM). *Inset:* Structural dependency of FtsZ-driven acoustic modes. LDV spectral amplitudes increase with FtsZ concentration but decrease with PA-matrix density.

**Figure 8C** shows the strain-dependent modulus $|\epsilon(\gamma)|$, capturing the nonlinear viscoelastic response of FtsZ–PA hydrogels under increasing deformation ($\gamma \equiv |u|$). In passive gels, the modulus remains constant across all strains, reflecting strain-invariant linear viscoelasticity in the absence of internal forces. In contrast, active gels display a marked strain-softening transition, with $|\epsilon(\gamma)|$ gradually decreasing and collapsing beyond a characteristic yield strain, $\gamma_Y \approx 1 - 10\%$, most prominently in softer networks (e.g., 3% PA). The transition occurs as the modulus crosses the threshold ($\epsilon = 1$), defining a "mechanical vacuum"—a bifurcation point beyond which the material offers no net elastic resistance and responds reactively to deformation. This marks the onset of mechanical depolarization, where extensile forces from FtsZ activity overpower the matrix's restoring stress, leading to extensile weakening ($|\epsilon(\gamma)| < 1$), characteristic of active soft matter. This strain-induced transition complements the **frequency-driven softening** observed in Fig. 8A–B: while $\epsilon'(\gamma < \gamma_Y)$ reflects dynamic compliance under fast internal fluctuations, $|\epsilon(\gamma > \gamma_Y)|$ reveals its manifestation under sustained strain. Together, they define two orthogonal bifurcation axes—frequency and strain amplitude, that govern the transition from passive stiffening to active softening. This strain-induced depolarization is quantitatively captured by the Landau phase transition model[29], where susceptibility $\chi \approx \pm \Gamma^2$ (order parameter) links mechanical polarization to internal activity, (see Supplementary Note N1). Equivalently, this activity connects with the effective temperature in a soft glassy rheology (SGR) theory $\chi \approx \mp \Delta x$ (SGR noise elevation; see Supplementary Note N2).

The integrated Landau-SGR model accurately reproduces the data across PA densities (see **Table I,** Supplementary Note SN2), showing that: a) More compliant, active gels require more negative susceptibiliy ($\chi < 0$; see Supplementary Note N1, Eqs. S4–S14); b) Yielding occurs at lower thresholds ($\epsilon_Y = |\epsilon(\gamma_Y)| < 1$; see Supplementary Note N1, Eqs. S15 and S19); c) Softer networks respond to activity over broader frequency ranges, consistent with the predictions of nonlinear rheological destabilization (see Supplementary Note N2, Eqs. S26–S32). These results confirm that strain-induced mechanical instability in active gels is a bifurcation phenomenon, driven by the interplay of network compliance and internal stress generation, precisely described by the SGR–Landau framework.

**Figure 8D** complements these rheological measurements with laser Doppler velocimetry (LDV) analysis of the acoustic spectrum in FtsZ–PA gels. Power spectral density (PSD) from a low-crosslink gel (3%) containing GTP-activated FtsZ filaments reveals distinct peaks at ~15 Hz, 17 Hz, 19.5 Hz and 23.5 Hz, within the 0–40 Hz acoustic band. These hydrodynamic modes, absent in both control PA gels and hybrid gels reinforced with passive FtsZ filaments, indicate the emergence of active acoustic resonances linked to internal treadmilling activity leading to softening in the presence of living polymers (see Supplementary Note N1, for instability regime and active mode amplification under extensile conditions). Underlying hydrodynamic modes arising from collective molecular motion establish a bottom-up connection to the frequency-dependent mechanical susceptibility, manifesting at the macroscopic level as modulated viscoelastic compliance in active FtsZ–PA gels. Together, these results confirm that FtsZ activity transforms the gel from a passively elastic material into a mechanically extensile metamaterial, capable of tuning its viscoelastic state via internal stress generation. The observed reduction of stiffness (via $\epsilon = 1 \pm \chi < 1$), viscous losses (via $\varphi \ll \chi$), strain-induced yielding ($\epsilon_Y \ll 1$), and the appearance of hydrodynamic modes in the acoustic spectrum (a signature of internal actuation), all point to a transition from passive viscoelastic resistance to active, dynamically compliant behavior. This internally driven softening reflects a shift from static elasticity to a fluid-like, dissipative regime, where filament activity effectively depolarizes mechanical resistance and injects rheological gain into the system. The frequency and strain dependence of this transition confirms the programmability of these biohybrid gels across the solid–fluid spectrum through internal filament dynamics.

**The hybrid FtsZ-PA gel is hyperfluid.** To investigate the dissipative properties of FtsZ–PA hybrid hydrogels, we analyzed the dependence of viscoelastic losses and flow resistance on both polymer concentration and temperature. These experiments were designed to probing how internal filament activity modulates the rheological energy landscape, particularly the activation barrier typically associated with thermally driven relaxation in passive networks. Particularly, concentration scaling revealed active suppression of viscous coupling. **Figure 9A** shows the scaling of the loss modulus with acrylamide concentration, $G''\sim c^\alpha$, for passive, monomeric, and active FtsZ–PA gels. In passive systems, $G''$ increased steeply with gel concentration, under scaling power-exponent $\alpha_{pass} = 6.1 \pm 0.2$, consistent with entanglement-controlled reptation under classical gelation behavior[38]. The addition of FtsZ monomers (without GTP promoting living polymerization) had negligible impact on this scaling. In contrast, active gels exhibited a significantly weaker scaling exponent, $\alpha_{act} = 2.8 \pm 0.3$, revealing a fluidization effect that grows more prominent at lower polymer densities. This deviation from passive scaling implies that internal activity from polymerizing FtsZ filaments disrupts viscous coupling across the meshwork, effectively renormalizing the dissipation mechanism. The inset of Figure 7A further illustrates this transition: at fixed PA concentration (3.5%), GTP addition triggers an abrupt drop in $G''$, reflecting the onset of active treadmilling and internal extensile stress generation.

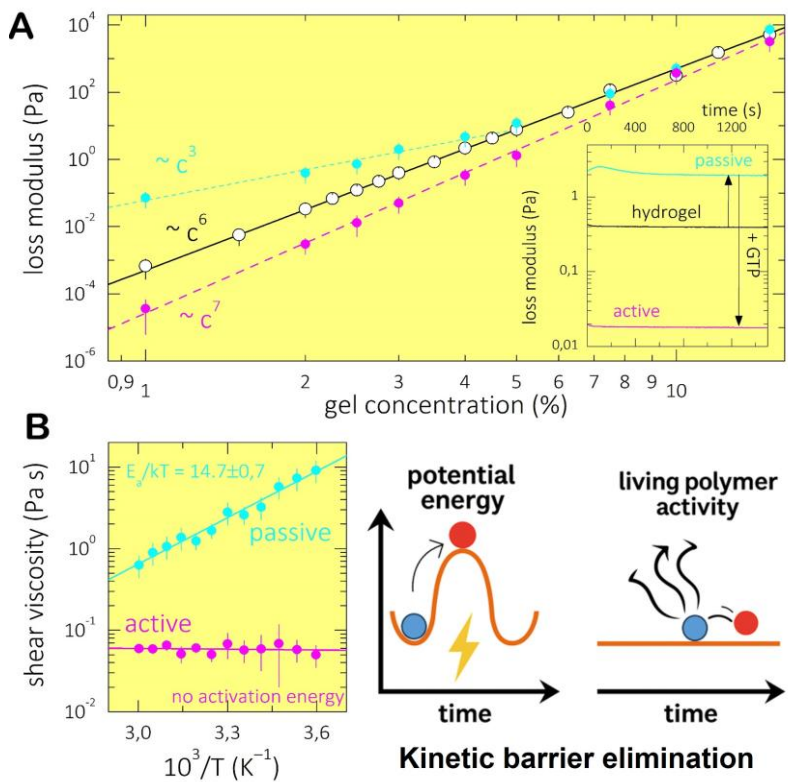

**Figure 9. Concentration dependence of viscous friction on the density network. A) Loss modulus** ($G''$) as a function of acrylamide concentration for passive gels (cyan), active hybrid gels (+GTP, magenta), and FtsZ-free controls (black). Passive PA matrices show a continuous scaling behavior, $G''_{pass}\sim c^6$, consistent with increasing frictional losses from the sol state up to dense network formation. Adding FtsZ filaments in the passive status increases further frictional losses, mainly in the sol state, providing a mark for passive hybridization, $G''_{pass}\sim c^3$. In contrast, active gels exhibit a much more intense scaling on concentration, $G''_{act}\sim c^7$, indicative of internal stress generation by living polymer activity that dominates over classical entanglement effects. Inset: time evolution of loss modulus at 3.5% PA, showing an abrupt drop upon GTP activation (arrow), reflecting internal fluidization. **B) Arrhenius plots of shear viscosity,** as $\eta = \eta_0 e^{-E_a/k_B T}$ at 3.5% PA (gel phase). Passive gels display a clear exponential dependence on inverse temperature, yielding an activation energy of $E_a/k_B T = 14.7 \pm 0.7$, consistent with thermally driven relaxation processes. In contrast, active gels show no measurable activation barrier, indicating temperature-independent internal dissipation driven by active FtsZ filament dynamics. This supports a nonthermal scenario in which active stress acts as an effective drift term, lowering the energy barriers for flow and mimicking a non-thermal internal driving force. ***Inset (bottom right):* Kinetic barrier elimination under filament activity** (the ball illustrates barrier-crossing in the active state driven by filament-induced softening): In the passive state (left), particle motion is confined by an energy barrier that must be thermally overcome. Upon FtsZ activation (right), internal stress and fluctuations from living polymers flatten the barrier, enabling effective drift-like flowing dynamics without requesting thermal activation.

**Temperature loss-independence confirms active elimination of flow barriers upon living polymerization.** To quantitatively assess these barriers, we measured the shear viscosity as a function of inverse absolute temperature $(1/T)$ under both passive and active conditions **(Figure 9B)**. Passive gels displayed a clear Arrhenius-type behavior:

$$\eta = \eta_0 e^{-E_a/k_B T}. \tag{4}$$

with an extracted activation energy $E_a/k_B T = 14.7 \pm 0.7$, consistent with values reported for crosslinked polymer networks where thermal fluctuations alone are insufficient to overcome the energy barrier for network rearrangement. This sets the baseline energy required to activate relaxation modes via external shear. By contrast, active gels exhibited a complete loss of temperature dependence over the same range, indicating a full collapse of the activation barrier. This result is striking given the estimated power injected by treadmilling FtsZ filaments, which can locally generate energy inputs on the order of $E_\xi \approx 10^3 k_B T \gg E_a$ (see Fig. 1). Such large non-thermal inputs drive the system far beyond equilibrium, effectively bypassing thermally activated pathways and enabling rearrangement through internally generated mechanical drift. The hybrid gel thus transitions into a hyperfluid regime characterized by active mechanical reactance ($\phi > 1, \epsilon < 1$), where flow is governed by endogenous stress generation rather than stochastic thermal agitation. These findings reveal a distinct nonequilibrium dissipation mechanism in active FtsZ-PA gels: rather than following conventional thermal soft-glassy rheology, they behave as internally driven, barrier-free fluids, whose mechanical impedance can be actively suppressed via biochemical control.

**DISCUSSION**

Differently from the processive living polymerization found in eukaryotic actin[53] and tubulin filaments[8]—which generate contractile stresses by pulling along persistent polar tracks, bacterial FtsZ filaments assemble into short, dynamic "living" structures that reorganize stochastically, acting as rotating force dipoles able to exert extensile stress away from their center-of-mass[16]. These living FtsZ polymers result in dilation and fluidization of the embedding hydrogel. The specific rheological response of the hybrid FtsZ–PA gel depends critically on two structural factors: (i) the tribological interaction between active filaments and the matrix cavities[¡Error! Marcador no definido.], which governs force transduction under mechanical contact; and (ii) the network connectivity[54], which controls the spatial coherence of internal stress propagation. The observed reductions in viscoelastic moduli, mechanical permitivitties, yield stress and the collapse of the Arrhenius activation barrier all point to the hyperfluid nature of the active gel, driven by internally generated forces rather than thermally activated flow. Soft active hydrogels formed by polymerizing protein filaments are common in nature[55], especially as functional components of cytoskeletal structures and extracellular matrices, where they contribute to tissue protection, mechanical signaling, and remodeling. The central concept demonstrated here involves embedding the bacterial self-assembling cytoskeletal protein FtsZ into a hydrogel polymer matrix. Upon GTP activation, FtsZ self-assembles and treadmills, converting soluble chemical energy into spatially localized motion that drives the material far from equilibrium. This functional building principle—incorporating self-assembling, force-generating proteins into compliant matrices—recapitulates design strategies found in natural soft matter systems that regulate processes such as cytokinetic fluidity control and self-healing in hydrogel biomaterial[56], muscular hydrostat function in animals[57], and reactive elasticity underlying nastic motions and tropisms in plants[58,59]. While the FtsZ protein provides a novel mechanism for generating extensile forces within a hydrogel network, the choice of polyacrylamide (PA) as the matrix material is also a critical design consideration. PA was initially selected for its biocompatibility, ease of functionalization, and well-characterized mechanical properties, allowing us to isolate and study the effects of FtsZ activity on the hybrid hydrogel's bulk behavior. Its relatively inert nature also minimizes unwanted interactions with the FtsZ protein, ensuring that the observed effects are primarily due to FtsZ-driven forces.

From this synthetic biology framework, we propose a novel class of active microgels, comprised of living FtsZ polymers embedded in PA matrices. These active hydrogels can be modularly assembled into higher-order structures with defined morphologies. The location and magnitude of stiffness gradients, along with the spatial distribution of internal FtsZ activity, collectively dictate the overall mechanical behavior. While PA offers advantages in initial characterization, its inherent limitations, including a lack of biodegradability and limited cell adhesion, necessitate exploration of alternative matrix materials. Future iterations could incorporate biodegradable polymers such as natural alginate, agarose and collagen, synthetically modified gelatin methacryloyl, among others, or enabling controlled scaffold degradation to facilitate tissue remodeling for FtsZ integration. These alternative biomaterials, although potentially presenting increased complexity in synthesis and processing, promise improved biocompatibility, biodegradability, and expanded functionality, ultimately broadening the range of applications for these active, extensile FtsZ-based hydrogels.

**Positioning Relative to Existing Active Gel Systems.** While established active gel systems based on actomyosin or microtubule-kinesin networks predominantly exhibit contractile behavior, leading to stiffening and bulk deformation, our FtsZ-based extensile gels invert this paradigm. These systems exhibit stress-induced softening and suppression of the Arrhenius flow barrier, offering a distinct mechanical profile. Unlike contractile systems, where motor activity reinforces matrix resistance and generates tension, the FtsZ-PA system uniquely demonstrates a reduction in viscoelastic moduli and yield stress upon activation. This extensile fluidization is achieved through the generation of internal pushing forces by polymerizing FtsZ filaments, disrupting the hydrogel network and promoting localized softening. In essence, the FtsZ gel mimics biological softening mechanisms utilized in tissue morphogenesis and cell migration, where localized matrix remodeling is crucial for cellular processes, rather than directly transmitting force across a macroscopic distance.

Further, while actomyosin-based systems excel at generating high forces and achieving significant shape changes, they often require precise control over filament organization and crosslinking. Microtubule-kinesin systems, known for their long-range transport capabilities, can be complex to implement and may exhibit limited force generation. The FtsZ-PA system, on the other hand, offers a relatively simple and robust approach to generating tunable extensile forces within a hydrogel. Although the force output may be lower than some contractile systems, the FtsZ system benefits from its bacterial origin, potentially enabling more cost-effective and scalable production. The inherent reversibility of FtsZ polymerization, coupled with its self-limiting behavior, also provides a level of dynamic control not readily achievable in other active gel systems. The reversible extensile fluidization exhibited by the FtsZ-PA system thus complements and expands the known motor toolbox, enhancing the functional landscape of active matter by providing a unique mechanical modality.

**Modularity Toward Functional Architectures.** The microgel-based formulation of hybrid materials enables a modular assembly strategy, where active units can be patterned or combined into higher-order architectures. This makes it feasible to construct materials with spatially resolved stiffness gradients, asymmetric response profiles, or localized activation zones i.e., functional features that are essential in soft machines and adaptive interfaces. Hence, we envision future embodiments where such active FtsZ–PA microgels are coupled to stimulus-responsive control layers, enabling programmable logic (e.g., gating, switching) based on mechanical input. The extensile mechanics of FtsZ filaments further lend themselves to non-reciprocal material behavior, in which deformation and relaxation follow distinct energetic pathways—a defining trait of living polymer materials.

## OUTLOOK

**Bacterial FtsZ Living Polymers: A Platform for Programmable Active Soft Matter.** The hybrid FtsZ–PA gels described here constitute a foundational platform for building internally programmable soft materials, in which viscoelastic response is not externally imposed but generated from within by biological activity. By tuning both the polymer network properties and the spatial distribution of active filaments, one can design materials that autonomously modulate stiffness, yield thresholds, and dissipation modes. This opens a path toward synthetic analogs of biological tissues, where dynamic mechanical adaptability is crucial; for instance, in wound-healing scaffolds, mechanoresponsive coatings, or flow-regulating membranes. Moreover, the reversible activation and self-limiting behavior of FtsZ filaments suggest potential for cyclical actuation or stress oscillation, laying groundwork for soft active robotic components that function without rigid motors or external pressure systems. The modular design of FtsZ-active microgels thus opens new avenues for creating biomimetic materials with muscular hydrostat-like responsiveness for use in tissue engineering, adaptive biomaterials and soft robotics.

## MATERIALS AND METHODS

**Preparation of FtsZ–Polyacrylamide Hybrid Hydrogels.** Recombinant *Escherichia coli* FtsZ protein was purified as described previously and stored at −80 °C in a GTP-free buffer. Hybrid hydrogels were prepared by co-polymerizing FtsZ within an acrylamide/bis-acrylamide network. The prepolymer solution contained acrylamide (2.5–10% w/v), N,N′-methylenebisacrylamide (0.07% w/v), 0.5 mg/mL FtsZ, and 1 mM GTP in a buffer composed of 50 mM HEPES, 100 mM KCl, pH 7.5. Polymerization was initiated by adding 0.05% ammonium persulfate (APS) and 0.05% TEMED under isotonic conditions. To prevent premature FtsZ activation, no $Mg^{2+}$ was present during polymerization. Gels were cast in flat or cylindrical geometries depending on the intended analysis (microscopy, rheology, or swelling assays).

**Activation and Control Conditions.** Gels were equilibrated overnight at 4 °C in the polymerization buffer (GTP included, $Mg^{2+}$ omitted). To activate FtsZ polymerization, samples were transferred to fresh buffer containing 2 mM $MgCl_2$ at 37 °C. Control experiments were conducted using (i) gels lacking GTP (monomeric FtsZ only), (ii) gels lacking FtsZ entirely (passive network), or (iii) non-activated conditions without $Mg^{2+}$.

**Microgel formation: swelling essay.** Microscale gel droplets (~100 μm diameter) were fabricated using a millifluidic injection setup composed of a fine-tipped glass capillary connected to a syringe pump. Droplets were polymerized directly into an isotonic aqueous bath containing acrylamide monomers, photoinitiators and cross-linkers mixed with FtsZ and GTP ($Mg^{2+}$ omitted), followed by overnight incubation to ensure completion of gelation and initial swelling equilibrium under hybrid polymerization. Bright-field microscopy (Nikon TE2000, 4× and 10× objectives) was used to image individual microgels after this first equilibration. FtsZ activation was then triggered by addition of $Mg^{2+}$, and the samples were further incubated until a second post-activation swelling equilibrium was reached after 1 hour. Quantification of radius and volume changes between the two equilibrium states (pre- and post-activation) was performed using image segmentation and circular contour fitting in ImageJ.

**Transmission Electron Microscopy (TEM).** Gels (2.5% and 7% PA) were fixed in 2% glutaraldehyde, post-fixed in 1% osmium tetroxide, dehydrated in ethanol, and embedded in epoxy resin. Ultrathin sections (70 nm) were stained with uranyl acetate and imaged using a JEOL JEM-1011 TEM at 100 kV to examine internal architecture and filament organization.

**Rheology Measurements.** Bulk rheological characterization was performed using a stress-controlled shear rheometer (TA Instruments) in a 40 mm cone plate configuration. Gels were equilibrated at 37 °C, and G′ (shear modulus) and η (viscosity) were measured in real time before and after $Mg^{2+}$ addition. Frequency sweeps (0.01–10 Hz) and strain sweeps (0.01–50%) were performed to assess viscoelastic behavior across the linear and nonlinear regimes. For Arrhenius analysis, viscosity was recorded at a constant shear rate (1 $s^{-1}$) across temperatures from 15 to 40 °C.

**Particle Tracking.** For Brownian motion tracking, hydrophilic microparticles (~1 μm diameter) were synthesized using an oil-in-water emulsion polymerization. A monomer mix composed of 60 mol% methyl methacrylate (MMA), 40 mol% poly(ethylene glycol) methacrylate (PEGMA, MW 300–500), and 0.5 mol% ethylene glycol dimethacrylate (EGDMA) was emulsified in 1% sodium dodecyl sulfate (SDS) via probe sonication. Polymerization was initiated with APS (0.5% w/v) and TEMED (0.05% v/v) under a nitrogen atmosphere at room temperature. The PEG-rich composition produced microparticles with a net density of 1.00 ± 0.01 $g/cm^3$, confirmed by sedimentation and sucrose gradient flotation assays, thus ensuring neutral buoyancy and preventing sedimentation during gel polymerization and measurement. The surface chemistry conferred by PEG also rendered the particles strongly hydrophilic and inert, minimizing non-specific adhesion or mechanical coupling to FtsZ filaments. After synthesis, particles were washed and resuspended in phosphate-buffered saline (PBS, pH 7.4) containing 0.01% Tween-20 to prevent aggregation. They were added to the FtsZ–PA pre-gel solution at a final concentration of ~$10^6$ particles/mL and gently mixed before casting to ensure homogeneous spatial distribution throughout the hydrogel matrix.

**Synthesis of Tracking Microparticles.** Carboxylated polystyrene microparticles were synthesized using a surfactant-stabilized dispersion polymerization in aqueous medium. The reaction mixture consisted of 10% (v/v) styrene monomer and 2% (w/v) acrylic acid as a comonomer to introduce surface carboxyl groups, emulsified in deionized water containing 0.5% sodium dodecyl sulfate (SDS). The emulsion was homogenized by magnetic stirring under nitrogen for 30 minutes to eliminate dissolved oxygen. Polymerization was initiated by addition of 0.5% ammonium persulfate (APS) at 70 °C under continuous stirring for 6 hours. The resulting colloidal suspension was cooled to room temperature and centrifuged at 5,000 g for 10 minutes. The particles were washed three times with deionized water and resuspended in phosphate-buffered saline (PBS, pH 7.4) containing 0.01% Tween-20 to prevent aggregation. Particle size and uniformity were confirmed by dynamic light scattering and optical microscopy, yielding a near monodisperse population of microparticles (1.05±0.03 µm, 0.103 PDI in diameter) with a surface density of carboxyl groups suitable for colloidal stability under physiological conditions. To ensure near-neutral buoyancy, the synthesis conditions were optimized to produce particles with a density close to water (1.03–1.05 g/cm³).

**Brownian Motion Analysis.** The hydrophilic, carboxylated surfaces of microparticle tracers minimized non-specific interactions with FtsZ filaments, enabling reliable tracking of thermally driven and active fluctuations when embedded in the hybrid FtsZ–PA hydrogel. To probe internal force fluctuations, 1-µm carboxylated polystyrene beads were embedded in the gel prepolymer solution prior to curing. Bead motion was recorded by high-speed video microscopy (100 fps) using a sCMOS camera (Photron) under 40× magnification in a phase contrast microscope (TE2000, Nikon). Trajectories were extracted using TrackMate (ImageJ), and displacement distributions, mean square displacements (MSD), and non-Gaussian parameters were computed to quantify deviation from equilibrium thermal diffusion.

**Laser Doppler Velocimetry (LDV).** To characterize the dynamic behavior of the FtsZ-PA hydrogels and quantify the FtsZ-driven hydrodynamic modes, we employed Laser Doppler Velocimetry (LDV). Measurements were performed using a LDV100 instrument (Polytec, Germany), equipped with a He-Ne (632.8 nm wavelenght) laser. The hydrogel sample was placed on the stage of the LDV instrument, and the laser beam was focused into the sample center using a focusing objective. The backscattered light from the hydrogel was collected by the same objective and directed to a photodetector. The Doppler frequency shift of the backscattered light, which is proportional to the velocity of the microparticles, was analyzed using a fast Fourier transform (FFT) to generate a power spectral density (PSD) spectrum.

**Data Analysis and Statistics.** All experiments were replicated at least three times. Data are presented as mean ± standard deviation unless otherwise stated. Curve fitting (power-law exponents, Arrhenius plots) and statistical analysis were performed using OriginPro and Python (SciPy). Significance of mechanical softening and volume changes was assessed using Student's t-tests, with $p < 0.05$ considered significant.

# ACKNOWLEDGEMENTS

We thank Germán Rivas and Miguel Vicente for providing the purified FtsZ samples, Iván López-Montero and Paolo Natale for assistance with hydrogels' preparation, and Marisela Vélez for discussions. The RSS of Instituto de Catálisis y Petroleoquímica (ICP)-CSIC, Spain is acknowledged for making SEM available for use. This research was supported by the Spanish Ministry of Science and Innovation (MICINN–Agencia Española de Investigación AEI) under grants FIS2015-70339-C2-1-R, PID2019-108391RB-100 and TED2021-132296B-C52 (to FM), TED2021-132296B-C53 (to DHA), and Comunidad de Madrid under grants S2018/NMT-4389 and Y2018/BIO-5207 (to FM). This study was also funded by the REACT-EU program PR38-21- 28 ANTICIPA-CM, a grant by Comunidad de Madrid and European Union under FEDER program, from European Union in response to COVID-19 pandemics. The funders had no role in the study design, data collection, analysis, preparation of the manuscript, or the decision to publish.